\documentclass{article}
\usepackage[utf8]{inputenc}   
\usepackage{amsmath}    
\usepackage{amsthm}     
\usepackage{amssymb}    
\usepackage{colortbl}  
\usepackage{comment}    
\usepackage{soul}       
\definecolor{mygrey}{rgb}{0.9, 0.9, 0.9}    
\usepackage{graphicx}   
\usepackage{parboxx}
\usepackage[affil-it]{authblk}  %

\newcommand{\mbf}{\mathbf}

\begin{document}

\title{ Fooled by bursts. A Goal per Minute model for the World Cup.  }
\author{Nicola Mingotti 
\thanks{Electronic address: \texttt{nicola.mingotti@uc3m.es} }}
\affil{Statistics Department, University Carlos III of Madrid, Spain}

\maketitle

\begin{abstract}
On the occasion of the last FIFA World Cup in Brazil, {\em The Economist}
published  a 
plot depicting how many goals have been scored in all World Cup competitions until present, 
minute by minute. The plot was followed by a naive and poorly grounded qualitative analysis. 
In the present article 
we use {\em The Economist} dataset to check its conclusions, update
previous results from literature and offer a new model. In particular, it will be shown 
that first and second half game have different scoring rates. In the first half 
the scoring rate can be considered constant. In the second it increases 
linearly with time.

\end{abstract}

\section{Introduction}
{\em The Economist} \cite{the.economist} is a well known weekly newspaper of economics and economics related
subjects. Frequently, it publishes interesting quantitative information in paper 
and on the web, especially in the {\em Daily chart} section. In occasion of the 2014 World Cup 
it collected the amount of goals scored 
during all the World Cup competitions, from 1930 to 2014, minute by minute. 
All the data is presented as interactive plot on the web \cite{gooaall}.

\begin{figure}[ht]
\centering
\includegraphics[scale=0.35]{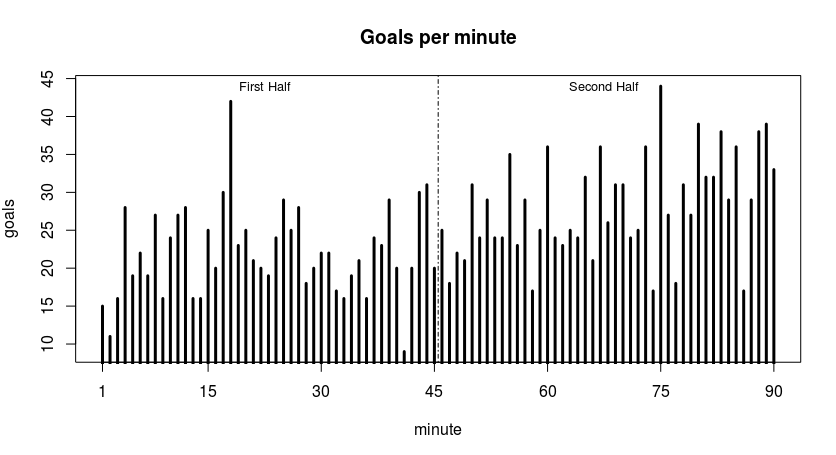}
\caption{Goals per minute. Reduction of {\em The Economist} original plot \cite{gooaall} displaying
all goals scored in all World Cup matches until July 2014, minute by minute. 
{\em Extra time} and {\em additional time} have been removed.  }
\label{fig:economist-reduction}
\end{figure}

We give in Fig.\ref{fig:economist-reduction} a reduction of {\em The Economist} original 
graphic. In our plot {\em additional time} (called {\em allowance for time lost} in 
FIFA documentation \cite{fifa7} ) and 
{\em extra time} are not represented.
No distinction is made between {\em goals}, {\em penalties} and {\em own goals}.
In the original graphic it possible to see in which match each goal was scored
and is possible to filter matches according to some categories. The reader 
interested in these facets should definitely refer to the web. All the data
used for this paper and the analysis performed in {\em R} \cite{r} software  are 
available at 
{\em mingotti.uc3m.es/floor1/paper-goals}.

Just under the original plot the authors summarize the graphical information with the following sentence:
{\em “... One can expect a rush of goals in the last ten minutes of normal time, but the 18th and 75th minutes have proved fertile”}. We consider such conclusions hurried and possibly lacking in 
significance. In this paper we will propose a simple and reasonable 
model that describes the minute by minute goal scoring frequency. 
After that, we reconsider how much of the original analysis can be 
confidently accepted.  

The goal scoring frequency has been already analyzed in literature for the 
{\em National Scottish League 1991-1992} in \cite{reilly1996}, the {\em Australian Soccer League} 
in \cite{abt2002} and the {\em European Champions League} in \cite{michailidis2004}. 
The 1986 World Cup in \cite{jinshan1986}, 1990 World Cup in \cite{jinshan1993} and jointly 1998, 
2002 and 2006 {\em World Cups} in \cite{armatas2007}. As 
far as we could establish, a comprehensive study about goal scoring frequency in 
all {\em World Cups} till present has never  been approached.

Article \cite{armatas2007} is the most similar to our work. It focuses on
World Cup scoring frequency and it considers more than a single tournament. 
The authors concluded that dividing the match in two 45-min parts, 
most goals were scored in the second half. 
Dividing the match in 15-min parts, most of the goals were scored in the last period (76-90) and 
there was a trend towards more goals scored as time progressed. In this paper
we will see all these conclusion can be confirmed and refined using 
the richer dataset of all World Cups scores.

\section{Analysis}
From the {\em The Economist} original data set we consider only goals scored in
{\em regular time}, that is in minutes between  $[1,45]$ and $[46,90]$. 
We ignore {\em additional time} because
its occurrence is decided match per match by the referee \cite{fifa7} and 
{\em extra time} because it depends on the scoring situation at the end of 
regular time. Their goal scoring rate can not be directly compared 
with {\em regular time}.

\begin{figure}
\centering
\includegraphics[scale=0.4]{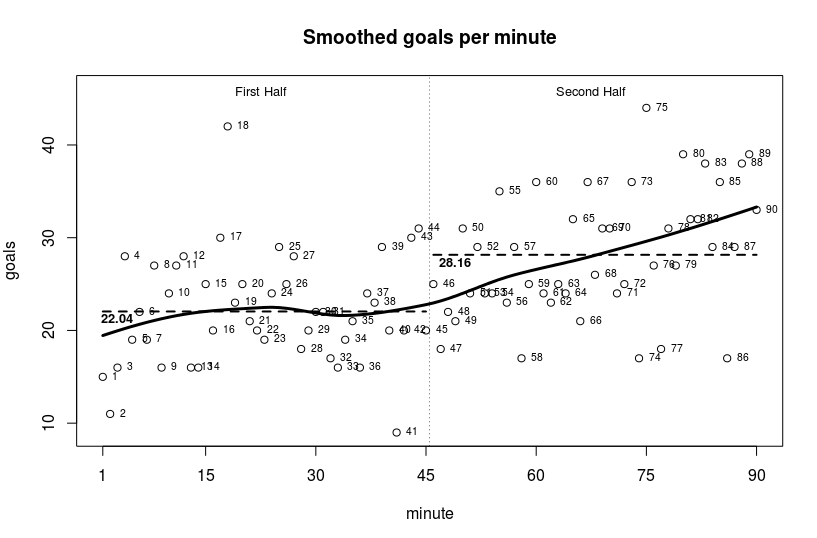}
\caption{Smoothed goals per minute. Goals per minute with a {\em loess}
smoother (thick line). Numbers near each circle represent  
minutes. For example, in the first half, the highest number of goals was scored on minute 18th,
the least on minute 41th. Dashed line represent half match goal averages.
The numerical value of each average is just under each dashed line, on its
leftmost side. 
}
\label{fig:loess}
\end{figure}

We redraw the dataset in Fig.\ref{fig:economist-reduction} 
with a {\em loess} smoother. In Fig.\ref{fig:loess} the 
thick continuous line is the smooth {\em loess} non parametric fit for the 
relation between goals and minute. The dashed lines represent the 
average number of goals in the first and second half respectively. Just under
the dashed lines, on their left side, do lay the numerical values of each mean.
The parameters used for the {\em loess} are the ones default with {\rm R} 
\cite{ venables2002, faraway2005, r}

From Fig.\ref{fig:loess} we can see the average number of goals in the second half (28.16)
is larger than in the first half (22.04). The difference between these 
averages $D := \bar{x}_2 - \bar{x}_1$  is statistically greater than zero.
In this work {\em significance level} is  always $\alpha = 0.05$.
Bootstrapping 10000 times we get $D$ is normally distributed as 
$N(\mu = 6.10, \sigma = 1.32)$.

Observing the {\em loess} smooth fit in Fig.\ref{fig:loess} we 
set the following hypothesis up. In the first half of the game 
the scoring rate is constant, in the second it grows 
linearly with time. We also notice that the first minutes could have 
an inferior scoring rate respect to the other minutes
in the first half. That is reasonable, the game starts always in the middle of 
the field, far from a comfortable scoring position. 

On the first half, our probabilistic model is that goals are
distributed according to a $\mbox{Multinomial}(n,(p_{1}, \dots, p_{45}))$ where
$p_{i} = p_{j}$ for all $i,j$ in $1 \dots 45$.
Performing a $\chi^2$ {\em goodness of fit} test 
we get $pval=0.0061$, homogeneity would be rejected.
But, if we remove just minute 18th, we get $pval=0.1296$ and homogeneity
would be far from rejected. We don't want to be too much lighthearted 
in dropping observations to satisfy our model so we reshape the dataset and
repeat the test. Instead of considering a match as composed
of a sequence of single minutes, we consider it 
as a sequence of 2,
3 and 5 minutes blocks\footnote{In order to have an even splitting 
of the dataset, when dividing a match in blocks of two minutes we ignore one observation,
minute 45th.}. For each minutes block 
 we set the associate number of goals as the average
number goals scored in each minute making the block. 
Performing again the $\chi^2$ {\em goodness of fit} test on the {\em first half},
now seen as sequence of 2,3,5 minutes blocks, we 
get {\em pvalue} respectively $(0.23, 0.49, 0.94)$. 
With a minimal smoothing reshape we see homogeneity
test in far from rejected, without dropping any observation.
In conclusion, we are confident that the {\em first half} can be considered
homogeneous and minute 18th an outlier. 


In the {\em second half} our working hypothesis is a linear
relation between minutes elapsed and goals scored. 
We begin fitting a simple linear regression model with {\em minute} as 
independent variable. By the {\em t-test}, {\em minute} is a significant
variable, $pval=0.00368$. Residuals can 
be considered normally distributed according to {\em Kolmogorov-Smirnov} and
{\em Shaipiro-Wilk} normality test, which give respectively
{\em pvalues} $(0.73, 0.60)$. $R^2 = 0.18$ is little and not 
much interesting for us. The fitted model appears in Eq.\ref{eq:model-1}
\begin{equation}
\label{eq:model-1}
goals = 13.346 + 0.0708 * minute 
\end{equation}

From Eq.\ref{eq:model-1} we can get the expected number of 
goals in minute $m \in (46,\dots,90)$. We will call these
values $\hat{\mbf{g}}_{2nd} := (\hat{g}_{46}, \dots, \hat{g}_{90})$.
Normalizing $\hat{\mbf{g}}$ we get 
$ {\hat{\mbf{p}}}_{2nd} := \hat{\mbf{g}}_{2nd} * (\sum_{i=46}^{90}{\hat{g_i}})^{-1} $.
And now we can apply again the $\chi^2$ to see if the model for
the {\em second half} is appropriate. Our null hypothesis is 
that goals realized in the second half ($\mbf{g}_{2nd}$) are
multinomially distributed with probability vector $\hat{\mbf{p}}_{2nd}$.
The $\chi^2$ test returns a pvalue of $0.09$ so the null hypothesis
can not be rejected in the second half game.

\begin{figure}[ht]
\centering
\includegraphics[scale=0.4]{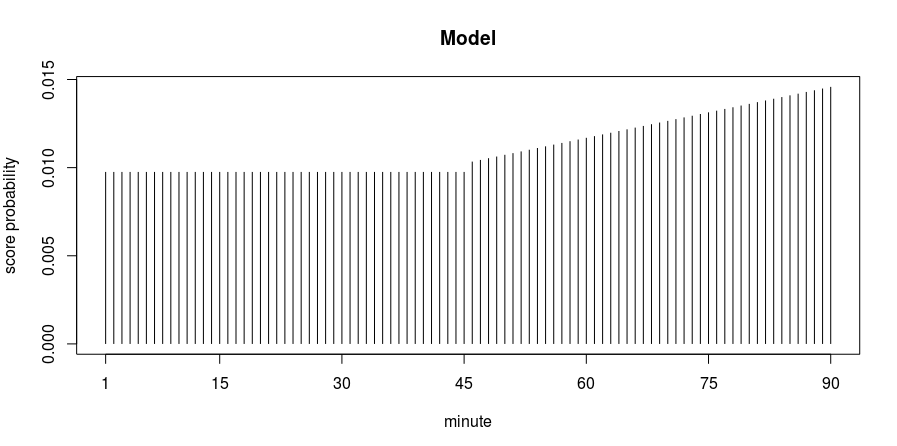}
\caption{The simple scoring model. If there is a goal in a match of the World Cup 
it will happen at some minute with the probability here depicted. In the first 
half game the probability is constant, in the second it grows linearly with time.}
\label{fig:model-scoring}
\end{figure}

\begin{figure}[ht]
\centering
\includegraphics[scale=0.4]{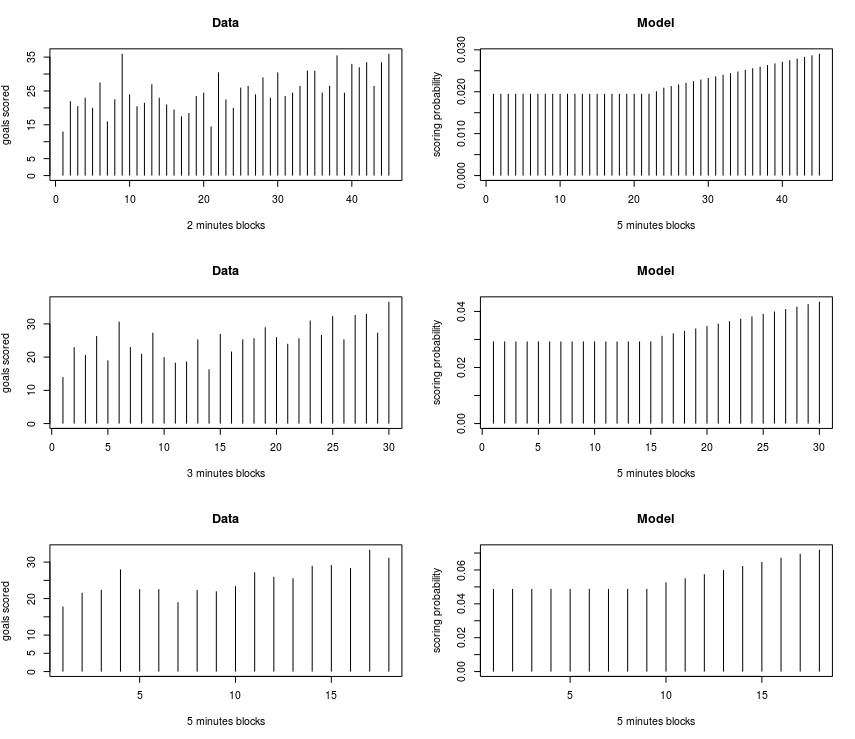}
\caption{This figure represents how reshaping variable {\em minute} in blocks of 2,3,5 units  
affects the goals distribution.}
\label{fig:model-per-blocks}
\end{figure}

Let us join the half time models and check hypotheses 
against the full length match data. According to our models,
the expected goal sequence is $(\tilde{g}_{1}, \dots, \tilde{g}_{90})$.
Where $\tilde{g}_{i}$ for $i \in [1,45]$ is the average number of 
goals scored in the first half and $\tilde{g}_{i}$ for $i \in [46, 90]$ 
is set by Eq.\ref{eq:model-1}. Normalizing $\mbf{\tilde{g}}$ we get 
a probabilities vector $\mbf{\tilde{p}}$, a graphical representation 
of it appears in Fig.\ref{fig:model-scoring}.
Using again a $\chi^2$, we test if 
goals scored in all Word Cups matches minute per minute
can be a realization of a {\em multinomial distribution} with probabilities vector $\mbf{\tilde{p}}$.
We get a $pvalue=0.0034$. Removing observation 18th we get 
$pval = 0.0524$ and we can not reject. Removing also the first three minutes
improves the fit giving $pval=0.1245$. To avoid removing
observation we reshape the dataset in blocks of 2,3,5 minutes
as done previously. The resulting {\em pvalues} are 
respectively $(0.86, 0.95, 0.99)$ so, the null hypothesis
is never rejected. Fig.\ref{fig:model-per-blocks}
illustrates goals smoothing on reshaping minute variable.

\begin{figure}[ht]
\centering
\includegraphics[scale=0.35]{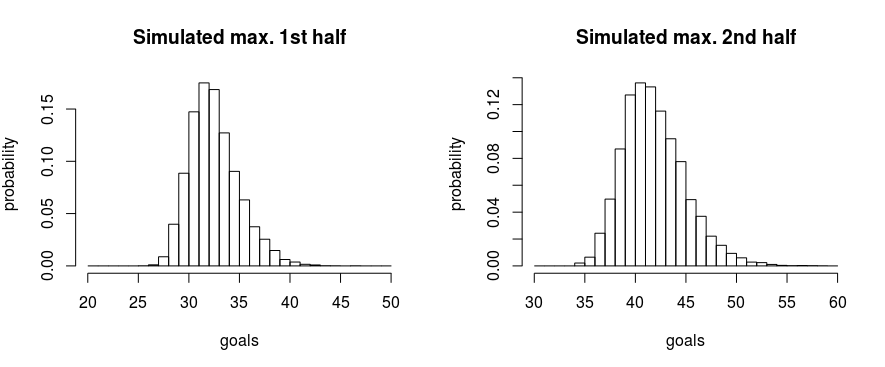}
\caption{Simulated maxima distribution for the first and second half.
These two histograms tell us that 42 goals, as a maximum in the first half,
is an unexpected result. On the contrary, 44 goals as a maximum for on the second half is  
not surprising.  }
\label{fig:simulated-max}
\end{figure}

An analytic description of the probability distribution can be easily
obtained once we move to a time continuous description.
Rescaling the match to a $[0,1]$ duration interval, 
if there will be a goal, the probability it will happen before
time $x$ will be given by $F(x)$. $F$ is defined in Eq.\ref{eq:cdf}.
\begin{equation}
\label{eq:cdf}
F(x) = \begin{cases}
0.878264\ x,  &  \quad 0 \le  x  < 0.5    \\
0.381889\ x^2 + 0.548938\ x + 0.069173,   &  \quad  0.5 \le x < 1 \\
1 &  \quad   x \ge 1  \\
\end{cases}
\end{equation}

Minutes 18th and 75th were called in the original document
``fertile'' minutes. Using our model we have a different interpretation
for each one of them.
In Fig.\ref{fig:simulated-max} we see the simulated distribution of the maximum\footnote{10000 maxima
 were simulated.} 
in the first and second half according to the model. 
For the first half we see that an observation
larger than the 18th (42 goals) is improbable, its probability is  
approximately 0.0015.
On the contrary, one observation larger than the 75th (44 goals) in the second
half it is not improbable at all ($p \approx 0.217 $).

\section{Conclusions}

There is enough evidence to state that the goal scoring
distribution in the World Cup matches can be considered 
constant in the first half game and growing linearly in the 
second half. Dividing the game in blocks of 2, 3 or 5 minutes 
confirms the model is valid and makes is more robust to 
extreme values. An analytic cumulative distribution function
for goal scoring is presented in Eq.\ref{eq:cdf}.

Previous finding in article \cite{armatas2007} are confirmed, 
there are more goals in the second half respect to 
the first one. According to our model, if there 
is a goal in a match, the probability that it will be in the 
first half is $44\%$.
The last part of the game is most probable for a goal,
and there is a trend toward more goals as time passes,
but only in the second half.

{\em The Economist} conclusions about its
dataset are only partially acceptable. Indeed, 
it is true that in on the last part of the game
there are expected more goals. It is not true
that minute 75th is an especially ``fertile'' minute, 
a maximum of 44 goals can be easily realized by
the variability of the second half game. 
Minute 18th is more interesting, according to our model
the probability of observing a maximum of 42 goals in the 
first half is about $2 \cdot 10^{-3}$. We guess it
has to be considered an outlier and removed because
it is an isolated burst, it has no connections
with its neighboring values. Performing a mild smooth,
as grouping each match as a sequence of 2 minutes 
removes all of its importance.


\bibliographystyle{plain}

\end{document}